\documentclass[10pt,letterpapers,floats]{article}
\usepackage{amsfonts,color,epsfig}
\usepackage{multirow}
\usepackage[multiple]{footmisc}
\usepackage{booktabs}
\newcommand{\nc}{\newcommand}

\newcommand{\rnc}{\renewcommand}
\renewcommand{\thefootnote}{\fnsymbol{footnote}}
\makeatletter
\rnc{\theequation}{\thesection.\arabic{equation}}
\@addtoreset{equation}{section}
\makeatother
\nc{\fig}[5]{ 
\begin{figure}[!htbp]
    \begin{center}
    \leavevmode
    \centerline{
        \includegraphics[width=#1, height=#2]{#3}
        }
    \caption[]{#4}
    \label{#5}
    \end{center}
\end{figure}}
\nc{\figs}[8]{
\begin{figure}[!htbp]
    \begin{center}
    \leavevmode
    \centerline{
        \includegraphics[width=#1, height=#2]{#3}
        \includegraphics[width=#4, height=#5]{#6}
        }
    \caption[]{#7}
    \label{#8}
    \end{center}
\end{figure}}

\begin{document}
\begin{flushright}
{\rm gr-qc/yymmnnn}
\end{flushright}
\vspace{14mm}
\begin{center}
{{{\Large {\bf Formation of Three-Dimensional Black Strings from Gravitational Collapse of Dust Cloud}}}}\\[10mm]
{Seungjoon Hyun$^{a,c}$\footnote{email:hyun@phya.yonsei.ac.kr},
Jaehoon Jeong$^{a}$\footnote{email:jejechi@gmail.com},
Wontae Kim$^{b,c}$\footnote{email:wtkim@sogang.ac.kr}, and 
John J. Oh$^{a}$\footnote{email:john5@yonsei.ac.kr}}\\[10mm]

{{${}^{a}$ Institute of Physics and Applied Physics, Yonsei University, Seoul, 120-749, Korea\\[0pt]       
${}^{b}$ Department of Physics, Sogang University, Seoul, 121-742, Korea\\[0pt]
${}^{c}$ Center for Quantum Space-time, Sogang University, Seoul, 121-742, Korea}\\[0pt]
}
\end{center}
\vspace{4mm}
\begin{abstract}
We study the formation of black strings from a gravitational collapse of cylindrical dust clouds in the three-dimensional low-energy string theory. New junction conditions for the dilaton as well as two junction conditions for metrics and extrinsic curvatures between both regions of the clouds are presented. As a result, it is found that the collapsing dust cloud always collapses to a black string within a finite collapse time, and then a curvature singularity formed at origin is cloaked by an event horizon. Moreover, it is also shown that the collapse process can form a naked singularity within finite time, regardless of the choice of initial data.  
\end{abstract}
\vspace{5mm}

{\footnotesize ~~~~PACS numbers: 04.20.Jb, 04.50.+h, 04.90.+e}


\vspace{2cm}

\hspace{11.5cm}{Typeset Using \LaTeX}
\newpage
\renewcommand{\thefootnote}{\arabic{footnote}}
\setcounter{footnote}{0}
\section{Introduction}\label{sec:intro}
One of important questions in general relativity is on the final fate of gravitationally collapsing compact objects such as massive stars. This question, to some extent, was answered by the {\it singularity theorems} in late sixties \cite{hp}, in which the gravitationally collapsing physically reasonable matter always gives rise to an emergence of spacetime singularities. However, they did not present any informations on their physical properties - for example, how about their energy densities, spacetime curvatures, and causal structures. The answers on these issues can be provided by means of the study of gravitational collapse in the context of general relativity.  The first theoretical study of collapsing homogeneous dust ball was achieved by  Oppenheimer and Snyder \cite{os}. They considered the interior metric describing the evolution of homogeneous dust as the Friedmann-Robertson-Walker (FRW) metric and matched it with the exterior Schwarzschild black hole solution on the edge of the dust cloud. As a result, they examined the dynamical evolution of spherically collapsing dust clouds and found that the collapse forms a Schwarzschild black hole as an outcome of the gravitational collapse. Many extensive studies on the gravitational collapse and a naked singularity have been done for more realistic case of inhomogeneous dust \cite{inho70,inho80,inho90}, matters with pressure \cite{press}, non-spherical symmetric matters \cite{lem}, and scalar fields \cite{scalar1,scalar2,scalar3}, which are based on the context of general relativity in four dimensions. 

On the other hand, after the discovery of the black hole solution in (2+1) dimensions \cite{btz}, the issue on the gravitational collapse became more intriguing in connection with the black hole formation and the naked singularity in anti-de Sitter (AdS) spacetimes. In the context of general relativity, several issues of the gravitational collapse have been extensively studied for a disk of pressureless dust ball \cite{mr}, a formation of conical singularities \cite{gak}, a black hole formation from colliding point particles \cite{mat}, the critical phenomena of black hole formation \cite{crit}, collapsing shell of radiation \cite{lemo}, and thin-shells of various sorts of matter including presureless dust, polytropic fluids (and perfect fluids), and the generalized Chaplygin gas (GCG) \cite{mo}, which are consistent with other results in four and two dimensions as well \cite{mr2}.
However, very little study on this issue in the framework of string theory has been done to date, despite it has been regarded as a promising candidate of quantum gravity. The central reason of this might be due to the difficulty of solving equations of motion with the dilaton and other charges. The gravitational collapse in two-dimensional dilaton gravity and matching conditions were investigated in ref. \cite{mr2}, which is the only study on the relevant issue in the viewpoint of the string theory. In this sense, it is interesting to study the issues on the gravitational collapse in the framework of string theory.

In this paper, the gravitationally collapsing cylindrical dust clouds in (2+1)-dimensional low-energy string theory is considered. As is well-known, there are two classes of solutions - one is the Banados-Teitelboim-Zanelli (BTZ) type black hole solution with a constant dilaton, another is the asymptotically flat black string solution with a nontrivial dilaton configuration, which are connected by $T$-dual symmetry \cite{hw}.
In order to deal with the collapse problem in string theory appropriately, apart from  the study in general relativity, one must require some additional matching conditions for the field contents in string theory (for example, the dilaton field and the Neveu-Schwarz (NS) two-form field), which will alter the equation of motion and determine the dynamical evolution of the collapsing matter. 

The main interest in our study is to see if the cylindrical pressureless dust collapses to either a black string solution or a naked singularity, comparing it with the results in AdS spacetimes and BTZ black holes \cite{mr,mo}. In section \ref{sec:exterior}, the black string solution is taken as an exterior spacetime and junction conditions on the edge regarding various fields in string theory are derived. In section \ref{sec:match}, this solution is matched to the exterior metric and exact solutions describing the evolution of the radius of edge are found. In section \ref{sec:interior}, the time-dependent collapsing solutions in the interior space are found on the edge.  In section \ref{sec:minkowski}, the collapse of dust clouds in Minkowski spacetimes is investigated.
Finally, we summarize and discuss our results in section \ref{sec:discussion}.

\section{The Exterior Black String Metric and Matching Conditions}\label{sec:exterior}

Let us start with the neutral sector of the three-dimensional low-energy string theory,
\begin{equation}
\label{eq:action}
S_{tot} = \frac{1}{2\kappa^2}\int dx^{3}\sqrt{-g}e^{-2\phi}\left[R+4(\nabla\phi)^2 + \frac{4}{\ell^{2}}\right] + S_{\rm M},
\end{equation}
where $\kappa^2 \equiv 8\pi G_{(3)}$, $\ell^{-2} = -\Lambda$ is a negative cosmological constant, and $S_{\rm M}$ is a matter action. Varying (\ref{eq:action}) yields equations of motion,
\begin{eqnarray}
&& e^{-2\phi}(R_{\mu\nu} + 2 \nabla_{\mu}\nabla_{\nu}\phi)=\kappa^2 T_{\mu\nu}^{\rm M},\label{eq:em1}\\
&& R - 4(\nabla\phi)^{2}+4\Box\phi + \frac{4}{\ell^{2}} = 0,\label{eq:em2}
\end{eqnarray}
where $T_{\mu\nu}^{\rm M} \equiv 2\delta S_{\rm M}/\sqrt{-g}\delta g^{\mu\nu} = \rho u_{\mu}u_{\nu}$.
Equations of motion (\ref{eq:em1}) and (\ref{eq:em2}) lead to a set of static uncharged black string solutions,
\begin{eqnarray}
&&(ds)^{2}=-{\mathcal F(R)}dT^{2} + \frac{\ell^{2}dR^{2}}{4R^{2}F(R)}+ dx^{2} , \label{eq:bsmetric}\\
&& \phi(R) = - \frac{1}{2}\ln(R\ell),
\end{eqnarray}
where
\begin{equation}
{\mathcal F}(R) = 1- \frac{\mathcal M}{R},
\end{equation}
which are dual to the three-dimensional BTZ black hole solution with the constant dilaton field \cite{hw}.
Note that $\{T,R,x\}$ is a coordinate system describing the exterior spacetime.

To derive junction conditions on the edge, let us define a distribution function as $\Theta(L) \equiv 1$ ($0$) for $L>0$ ($L<0$), where $L$ is a geodesic coordinate. This distribution function has the following properties,
\begin{equation}
\label{eq:prop}
\Theta(L)\Theta(-L) = 0, ~~\Theta^2(\pm L) = \Theta(\pm L),~~ \frac{d}{dL}\Theta(L) = \delta(L),
\end{equation}
where $\delta(L)$ is a Dirac's delta function. Decomposing the metric and the dilaton, $g_{\mu\nu} = \Theta(L) g_{\mu\nu}^{+} + \Theta(-L) g_{\mu\nu}^{-}$ and $\phi = \Theta(L)\phi^{+} + \Theta(-L)\phi^{-}$, where $(+)$ and $(-)$ denote the outer and the inner spacetimes, respectively, then the first junction conditions from the continuity of their first derivatives are found,
\begin{equation}
\label{eq:1stjc}
[g_{ij}] = 0, ~~[\phi] = 0,
\end{equation}
where $[A]\equiv A^{+} - A^{-}$. The second derivatives of the metric and the dilaton field will produce the second junction conditions . Since the second derivative of the metric depends on the extrinsic curvature along the hypersurface of the edge, the second junction condition becomes
\begin{equation}
\label{eq:2ndjc}
[K_{ij}]=0,
\end{equation}
while the second junction condition for the dilaton field {to the tangential direction} identically vanishes since its first derivative
is normal to the hypersurface.{\footnote{See the chapter 3 in ref. \cite{poi} for the explicit derivation.}\footnote{Adding NS-field needs an additional junction condition, $[B_{ij}]=0$ and the solutions for NS field between inside and outside should be matched. This requires that the only non-vanishing component of the NS-field inside should be $B_{xt}$ and furthermore it depends only on the time $t$, i.e. $B_{xt}=B_{xt}(t)$. Due to the antisymmetricity, the field strength for this configuration vanishes inside the clouds.  See ref. \cite{ckks} for cosmology of antisymmetric tensor fields on D-brane.}} {However,
the condition to the normal direction is given by
\begin{equation}
[n^{\alpha} \partial_{\alpha}\phi] = 0, \label{eq:2ndjc}
\end{equation}
in the absence of the dilaton source on the hypersurface.}\footnote{Detailed analyses of the singular hypersurfaces for the scalar-tensor theories of gravity are shown in ref. \cite{bb}}

First of all, we need to investigate the junction condition of the dilaton to the normal direction, eq. (\ref{eq:2ndjc}), which is expressed in the form of
\begin{equation}
\frac{2}{\ell} F \dot{T} = n^{r}\partial_{r}\phi,
\end{equation}
where $n^{\alpha}$ is a normal vector to the hypersurface.
Since there exists a nontrivial dilaton radiation toward the normal direction \cite{bb}, the inner configuration of the dilaton field cannot have a homogeneous one such as $\phi=\phi(t)$, instead it should be at least $\phi=\phi(t,r)$. Otherwise, $\dot{T}=0$ leads to a strong restriction to the equation of motion, which admits a solution, starting to collapse only from the inside an event horizon and this is not what we wish. 

In general, the collapse of dust ball with a non-trivial dilaton field is expected to have a time-dependent exterior solution since there obviously exists a dilaton radiation as it collapses. This can be obtained by assuming the homogeneous dust distribution inside the cloud. However, we wish to see the gravitational collapse of dust to the static black string solution as a final state of the collapse. In this case, the inhomogeneous setup inside the cloud is inevitable.
In addition, the interior metric should have the same topological structure of ${\bf R^2}\otimes{\bf S^{1}}$ to the exterior metric, eq. (\ref{eq:bsmetric}) since they should be smoothly matched on the edge of the cloud. From these facts, the interior space of inhomogeneous dust clouds is chosen to be the cylindrical form of the three-dimensional metric, comparing to the exterior metric, eq. (\ref{eq:bsmetric}),
\begin{equation}
\label{eq:inmet}
(ds)^2 = - dt^2 + a^2(t,r)dr^2 + dx^2,
\end{equation}
where $a(t,r)$ is a scale factor and $\{t,r,x\}$ is a comoving coordinate system. 

\section{Matching on the Edge}\label{sec:match}

In the exterior coordinates, the energy-momentum tensor, $T_{\mu\nu}^{\rm M}$, vanishes and
the exterior metric is governed by the metric (\ref{eq:bsmetric}). In the interior metric, the edge of the dust
cloud is located at $r=r_{0}$ while it is located at $R={\mathcal
  R}(t)$ in the exterior metric. Then the exterior metric on the edge of the cloud is 
\begin{equation}
\label{eq:bsmetricedge}
(ds)^{2}_{\rm edge} = - {\mathcal F}({\mathcal R}) dT^{2} + \frac{d{\mathcal R}^{2}}{{\mathcal F}({\mathcal R})\Omega({\mathcal R})} + dx^{2},
\end{equation}  
where $\Omega({\mathcal R}) \equiv 4{\mathcal R}^2/\ell^2$.
Since the boundary on the edge should be connected
smoothly, matching conditions for metrics and dilaton (\ref{eq:1stjc}) and (\ref{eq:2ndjc}) will be required \cite{isr,chase}. Then, the induced metric on the edge at $r=r_0$ is
\begin{equation}
  \label{eq:edgemet}
  (ds)_{\rm edge}^2 = - dt^2 + dx^2,
\end{equation}
and the first junction condition $[g_{ij}]=0$ gives
\begin{equation}
  \label{eq:metricmatch}
  \dot{T} = \frac{1}{\mathcal F} \sqrt{\frac{\dot{\mathcal R}^2}{\Omega}+{\mathcal F}}
\end{equation}
for the successful matching of
both metrics.

On the other hand, the extrinsic curvature tensor is defined by
\begin{equation}
  \label{eq:extcurv}
  K_{ij} = - n_{\sigma}\frac{\delta e_{(i)}^{\sigma}}{\delta \xi^{j}}
  = - n_{\sigma}(\partial_{j} e_{(i)}^{\sigma} +
  \Gamma_{\mu\nu}^{\sigma} e_{(i)}^{\mu}e_{(j)}^{\nu}),
\end{equation}
where $n_{\sigma}$ is the normal to the edge of the dust clouds,
$e_{(i)}^{\sigma}$ are the basis vectors on the edge, and $\xi^{j}$
refer to the coordinates on the edge.
In the interior coordinates, we have basis vectors,
\begin{equation}
  \label{eq:basisin}
  e_{(0)}^{\sigma}=(1,0,0),~~e_{(1)}^{\sigma}=\left(0,0,\frac{1}{a(t)}\right),
\end{equation}
and the unit normal vector, $n_{\sigma}=\left(0,a(t),0\right)$ while in the exterior coordinates, one finds
\begin{equation}
  \label{eq:basisout}
  e_{(0)}^{\sigma}=({\dot{T}},\dot{\mathcal R},0), ~~e_{(1)}^{\sigma} =
  \left(0,0,1\right),
\end{equation}
and $n_{\sigma}=\left(-{\dot{\mathcal R}}/{\sqrt{\Omega}},{\dot{T}}/{\sqrt{\Omega}},0\right)$.
It is found that all components of the extrinsic curvature tensor vanish in the interior coordinates
while the nonvanishing
component of the extrinsic curvature tensor in the exterior
coordinates is given by
\begin{equation}
\label{eq:ecvp}
K_{tt}^{+} = -\sqrt{\Omega}\frac{d}{d{\mathcal R}} \sqrt{\frac{\dot{\mathcal R}^2}{\Omega}+{\mathcal F}}.
\end{equation}
Thus the second matching condition for the extrinsic curvature, $[K_{ij}]=0$, leads to a equation of motion for ${\mathcal R}(t)$,
\begin{equation}
\frac{d}{d\mathcal R} \sqrt{\frac{\dot{\mathcal R}^2}{\Omega} + {\mathcal F}}  = 0, \label{eq:jeq1}
\end{equation}
which has a solution of
\begin{equation}
\frac{\dot{\mathcal R}^2}{\Omega} + F = \eta^2,
\end{equation}
where $\eta$ is an integration constant.

Alternatively, eq. (\ref{eq:jeq1}) can be rewritten in the form of an effective potential as
\begin{equation}
\label{eq:eomR}
\dot{\mathcal R}^2 + V_{\rm eff}({\mathcal R}) = 0,
\end{equation}
where
\begin{equation}
\label{eq:effpt}
V_{\rm eff}(\mathcal R) = \Omega ({\mathcal F}-\eta^2) = \frac{4{\mathcal R}^2}{\ell^2} \left(1- \eta^2-\frac{\mathcal M}{\mathcal R}\right).
\end{equation}
Note that ${\eta}$ is an integration constant determined by an initial condition as
\begin{equation}
\eta^2 = 1 - \frac{\mathcal M}{{\mathcal R}_0} + \frac{u_{0}^2 \ell^2}{4{\mathcal R}_{0}^2},
\end{equation}
where $u_{0} \equiv \dot{\mathcal R}_{0}$ is an initial velocity of the cloud edge and ${\mathcal R}_{0}$ is an initial position of the edge of the dust cloud. The plot of the effective potential is depicted in Fig. \ref{fig:effbs}.

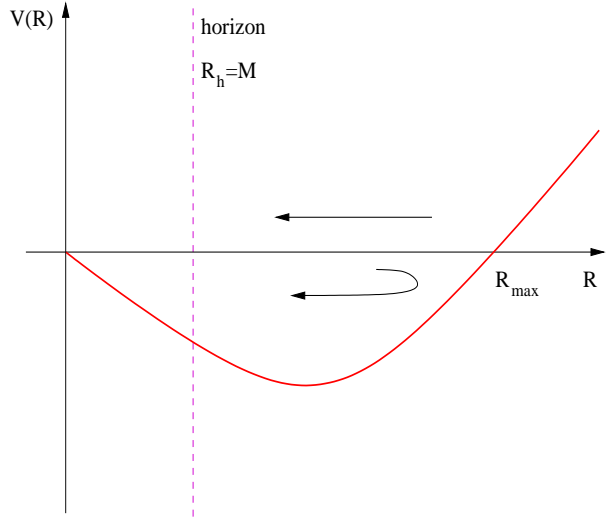
\fig{8cm}{7cm}{effbs}{\small A cartoon view of the effective potential for the formation of the black string. Once the initial position is fixed, ${\mathcal R}_{0}<{\mathcal R}_{\rm max} = {\mathcal R}_{0} + \frac{u_{0}^2\ell^2}{4{\mathcal M}{\mathcal R}_0}$ unless $u_{0} = 0$, then the edge of dust clouds collapses to a black string.}{fig:effbs}

An exact solution of eq. (\ref{eq:eomR}) is
\begin{equation}
\label{eq:Rsol}
{\mathcal R}(t) = \frac{\mathcal M}{\eta^2-1} {\rm sinh}^2\left[\frac{\sqrt{\eta^2-1}}{\ell} (t-t_{c})\right],
\end{equation}
where the collapse time $t_c$ is found to be
\begin{equation}
\label{eq:colt}
t_{c} = \frac{\ell}{\sqrt{\eta^2-1}} {\rm arcsinh}\sqrt{\frac{(\eta^2-1){\mathcal R}_0}{\mathcal M}},
\end{equation}
which is clearly finite and positive for all allowed ${\mathcal R}_0$ and $u_{0}$.
The behavior of collapsing dust edge is depicted for varying black string masses and initial velocities in Fig. \ref{fig:soletan1}.
One of interesting properties of this solution is that the final velocity of the edge, $\dot{\mathcal R}_c\equiv \dot{\mathcal R}(t_c)$ vanishes, which implies that the dust edge will stop at the end of the collapse. This is drastically different from the results in refs. \cite{mr,mo}.

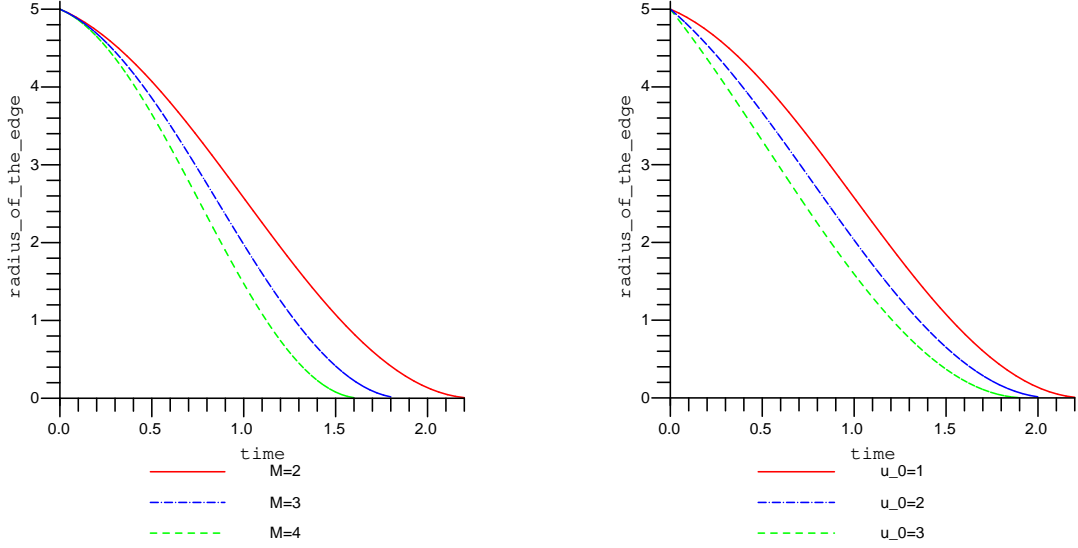
\figs{8cm}{8cm}{soletan1}{8cm}{8cm}{soletan1u}{\small a) LHS: Plots of the collapsing dust egde for various values of $\mathcal M$ for $u_0=1$ when $\ell=1$ and ${\mathcal R}_0=5$. b) RHS: Plots of the collapsing dust edge for various values of $u_0$ for ${\mathcal M}=2$ when $\ell=1$ and ${\mathcal R}_0=5$.}{fig:soletan1}

On the other hand, an event horizon will form around the collapsing dust at ${\mathcal R}_{h}$ and the comoving time $t_{h}$ at which
the event horizon and the dust edge coincide can be found at a
position of an event horizon, ${\mathcal R}_{h}$,

\begin{equation}
  \label{eq:comvtime}
  t_{h} = t_{c} - \frac{\ell}{\sqrt{\eta^2-1}} {\rm arcsinh}\sqrt{\frac{\eta^2-1}{\mathcal M} {\mathcal R}_{h}},
\end{equation}
which is clearly finite and positive. However, the coordinate
time at which an observer outside the dust clouds will observe the
collapse differs from the comoving time. 
A light signal coming from the dust surface at $T_f$ should satisfy the
null condition, $dR/dT=2(R-{\mathcal M})/\ell$ and it arrives at $R_f$
at time,
\begin{equation}
  T_f= T_0 + \int_{R_0}^{R_f} \frac{\ell dR}{2(R-{\mathcal M})} = T_{0} + \frac{\ell}{2} \ln\left(\frac{R_f -{\mathcal M}}{R_0 -{\mathcal M}}\right),
\end{equation}
which clearly diverges as $t\rightarrow t_{h}$ ($R_f \rightarrow {\mathcal M} = R_{h}$).
Therefore, the collapse of dust cloud is unobservable from the outside observer when it coincides with the black string horizon.

The ratio of the comoving time interval $dt$ between wave crests from
the dust and the interval $d\tilde{T}$ between arrived wave crests to
observers is equal to that of the natural wavelength $\lambda$ emitted
with no gravitation and the observed wavelength $\tilde{\lambda}$,
which defines the redshift from the dust edge by
\begin{equation}
  \label{eq:redshift}
  z=\frac{\tilde{\lambda}}{\lambda}-1 = \frac{d\tilde{T}}{dt} - 1 = \sqrt{\frac{\ell^2\dot{\mathcal R}^2}{4({\mathcal R}-{\mathcal M})^2} + 1 } - \frac{\ell\dot{\mathcal R}}{2({\mathcal R}-{\mathcal M})} - 1.
\end{equation}
The redshift diverges when $t$ approaches to $t_{h}$ (i.e. ${\mathcal R} \rightarrow {\mathcal M}={\mathcal R}_{h}$),
which implies that the dust cloud fades from the observer's
eyesight.

We note that the only spacetime singularity is at ${\mathcal R}=0$, which is easily seen by evaluating the Kretschmann scalars, ${\mathcal K} = {16{\mathcal M}^2}/{R^2\ell^4}$ in the exterior coordinates and ${\mathcal K} = {4}\left[ 2\ddot{a}^2 + \dot{a}^4\right]/a^4$ in the interior coordinates.

\section{The Interior Collapsing Dust Solutions}\label{sec:interior}

In the interior spacetime, the Eintein equations given by the use of the interior metric ansatz, eq. (\ref{eq:inmet}), are
\begin{eqnarray}
&& e^{-2\phi} \left(-\frac{2\ddot{a}}{a} + 2\ddot{\phi}\right) = \kappa^2 \rho,\label{eq:ine1}\\
&& a\ddot{a} + 2\left(\phi''-a\dot{a}\dot{\phi}-\frac{a'}{a}\phi'\right) = 0,\label{eq:ine2}\\
&& a^2\dot{\phi}^2 - \phi'^2 - a^2\ddot{\phi}+\frac{a^2}{\ell^2} = 0,\label{eq:ine3}
\end{eqnarray}
where the first two equations are the ones from the variation with respect to the metric while the last one is the dilaton equations. The conservation of the energy-momentum tensor, $\nabla_{\mu}T^{\mu\nu}=0$, yields $\rho(t,r)a(t,r)=\rho_0(r)a_0(r)$.
 The matching conditions for the dilaton, $[\phi]=0$ and $[n^{\alpha}\partial_{\alpha}\phi]=0$, yield 
\begin{eqnarray}
&&e^{-2\phi_e(t,r_0)} = {\mathcal R}(t)\ell \label{eq:2ndjcdil}\\
&&\phi'_{e}(t,r_0) = -\frac{a_e}{2\mathcal R} \sqrt{\dot{\mathcal R}^2 + \frac{4{\mathcal R}^2}{\ell^2}\left(1-\frac{M}{\mathcal R}\right)}, \label{eq:2ndjcdil2}
\end{eqnarray}
where $\phi_e\equiv \phi_e(t,r_0)$.
Plugging these into eq. (\ref{eq:ine3}), one finds
\begin{equation}
\dot{\mathcal R}^2 - {\mathcal R}\ddot{\mathcal R} - \frac{2M{\mathcal R}}{\ell^2} = 0,
\end{equation}
which is equivalent to eq. (\ref{eq:eomR}). This implies that the combination of the second junction condition and the dilaton equation are nothing but the equation of motion from the second junction condition with respect to the extrinsic curvature. Therefore, the second junction condition for the dilaton gives no additional constraint to the equation of motion, (\ref{eq:eomR}).

Eqs. (\ref{eq:ine3}), (\ref{eq:2ndjcdil}), and (\ref{eq:2ndjcdil2}) yield a solution for ${\mathcal R}(t)$ on the edge at $r=r_0$, which will be used as a boundary condition in solving the interior equation of motion. The remaining equations are eqs. (\ref{eq:ine1}), (\ref{eq:ine2}), and the energy-momentum conservation law as seen above. Generically, they cannot be solved because of the deficiency of the number of equations since there are six unknowns but only three independent equations. However, on the edge at $r=r_0$, eqs. (\ref{eq:2ndjcdil}) and (\ref{eq:2ndjcdil2}) can be used as a boundary condition and we already know the edge solution of ${\mathcal R}(t)$.
More precisely, substituting eq. (\ref{eq:2ndjcdil}) into eq. (\ref{eq:ine1}) on the edge, $r=r_0$ yields
\begin{eqnarray}
{\mathcal R}\ell \left(-\frac{2\ddot{a}_e}{a_e} + \frac{\dot{\mathcal R}^2}{\mathcal R^2} - \frac{\ddot{\mathcal R}}{\mathcal R}\right) = \kappa^2 \rho_e = \kappa^2 \frac{\rho_0a_0}{a_e}, \label{eq:eqnedge}
\end{eqnarray}
where $a_e \equiv a(t,r_0)$. Using eqs. (\ref{eq:Rsol}), eq. (\ref{eq:eqnedge}) can be solved by setting some parameters in a numerical manner.\footnote{The fourth-order Runge-Kutta method was used here.} Indeed, we used a collapsing initial condition, $\dot{a}<0$ and $\dot{\mathcal R}<0$, in solving the equations of motion.  
It is easily checked that the scale factor $a^{e}(t)$ vanishes at the collapse time, eq. (\ref{eq:colt}), by setting these initial conditions appropriately. In general, one finds a certain relation between parameters inside and outside the cloud such as $\rho_0$ and $M$, which might be obtained by identifying the collapse time of the edge, eq. (\ref{eq:colt}) with the collapse time at which the scale factor vanishes. However, in this analysis, it is not easy to find the precise relation between those initial conditions since we do not know the analytic solution inside the cloud. Instead some numerical searches find that there exists the relation by fixing some initial conditions.

Numerical plots for the scale factor, the energy density, and the dilaton field are illustrated in Fig. \ref{fig:scalerho}.

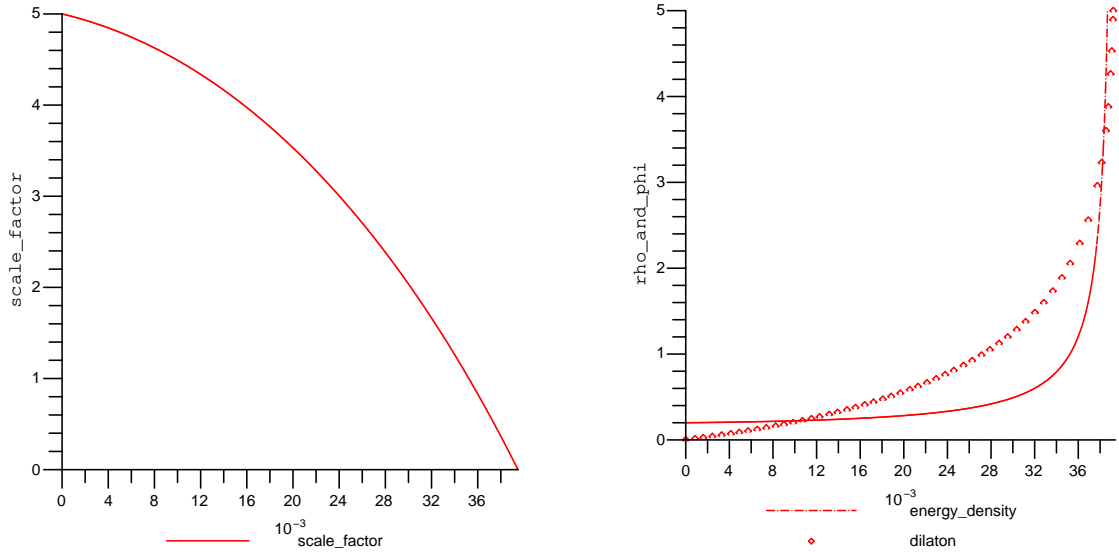
\figs{8cm}{8cm}{scalerhoinner}{8cm}{8cm}{densdilaton}{Some numerical solutions of the dust cloud in the interior spacetimes: the lhs describes a collapsing dust cloud solution while the rhs shows the diverging energy density and dilaton field. We set parameters for this numerical analysis as $\kappa=100$, $\rho_0=1$, $\ell=1$, $\mathcal R_0=5$, $a_0^{e}=5$, $M=952$, $\dot{a}_0^{e}=-30$, and $\dot{\mathcal R}_0=-30$.}{fig:scalerho}

\section{Collapse in Minkowski Spacetimes}
\label{sec:minkowski}

The main interest of this section is to see if the collapse can form a naked singularity in a finite collapse time. The exterior metric is assumed to be the Minkowskian metric, ${\mathcal F} = 1$, which is equivalent to ${\mathcal M} = 0$.\footnote{Rigorously speaking, the spacetime is the linear dilaton vacuum (LDV) since there still exists a dilaton solution, eventhough the metric solution describes Minkowski spacetimes.} Then the equation of motion is written as
\begin{equation}
\dot{\mathcal R}^2 + V_{\rm eff}({\mathcal R}) = 0,
\end{equation}
where the effective potential is $V_{\rm eff} (\mathcal R) = - u_0^2{\mathcal R}^2/{\mathcal R}_0^2$. Because of the shape of the potential (see Fig. \ref{fig:effms}), there are two sorts of collapse scenarios, depending on its initial velocity, $u_0$.

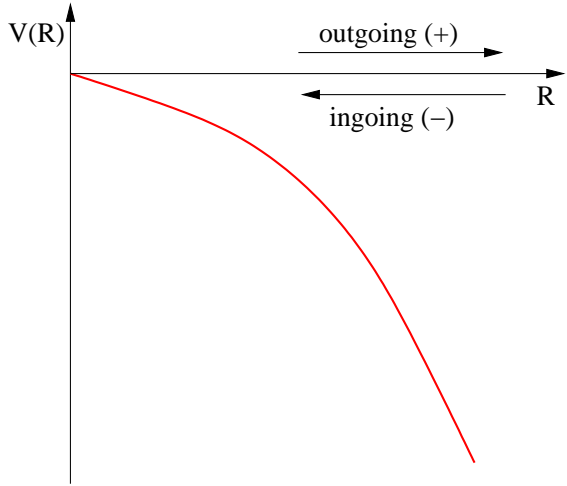
\fig{7.5cm}{6.5cm}{effms}{\small A cartoon view of the effective potential when the exterior spacetime is Minkowskian. The collapse scenarios depend on the initial velocity of the edge.}{fig:effms}

An exact solution is evaluated by ${\mathcal R}(t) = {\mathcal R}_0 e^{\pm |u_0| t/{\mathcal R}_0}$, which shows that if the positive sign is chosen, then the edge will expand indefinitely while the edge will collapse to a point for the negative sign. This feature is shown in the Fig. \ref{fig:effms}. At this stage, one may have a question on the final state after the collapse. Since there is no even horizon, if the dust cloud collapses to a point and forms a curvature singularity, then a naked singularity will appear. 
Indeed, the Kretschmann scalar ${\mathcal K}$ (or Ricci curvature scalar) diverges at the origin. However, the collapse time that ${\mathcal R}(t_c)=0$ also diverges, which implies that the comoving observer never meet a curvature singularity at the origin within finite time. This feature is also shown in the limit of ${\mathcal M}=0$ in eq. (\ref{eq:colt}). However, the preceding discussion is the case in string frame. Taking into account a transformation,
$g_{\mu\nu} \rightarrow e^{4\phi} g_{\mu\nu}^{\rm E}$, where $g_{\rm E}$ is an Einstein metric, the collapse time in the Einstein frame is \begin{equation}
\tau_c = \int_0^{t_c} e^{-2\phi}dt = \int_{0}^{t_c} {\mathcal R}(t)\ell dt = \frac{{\mathcal R}_0^2\ell}{|u_0|} \left(1- {|u_0|}e^{-\frac{|u_0|}{{\mathcal R}_0}t_c}\right).
\end{equation}
Since the collapse time in string frame, $t_c$, is infinite, one finds $\tau_c = {\mathcal R}_0^2\ell/|u_0|$. Notice that except for $u_0=0$, the collapse time is clearly finite.

We, therefore, conclude that a naked singularity appears from the collapse of dust clouds in Minkowski spacetimes (or LDV) in Einstein frame, which agrees to the previous results in refs. \cite{mr,mo}.

\section{Discussions}\label{sec:discussion}
We have demonstrated the formation of (2+1)-dimensional black string from the gravitationally collapsing cylindrical dust ball in the viewpoint of the low-energy string theory.
Provided the dust clouds collapse and form an event horizon at a certain position, the curvature singularity at ${\mathcal R}=0$ - rigorously speaking, a singular string at the origin along $x$-direction - is cloaked by an event horizon and  the dust cloud collapses to a three-dimensional black string. 
Once an initial position of the edge of dust clouds is fixed, the effective potential has a negative value within a region, ${\mathcal R}_{\rm max}$, which gives rise to a contraction of the edge, regardless of its initial velocity and the edge will collapse to a point at $t_c$, at which the dilaton field and the energy density diverge, respectively. However, unlike the previous works \cite{mr,mo}, the edge (or the scale factor) has a zero velocity at $t=t_c$, which implies that the edge will stop contracting after the collapse, where a black string along $x$-direction forms. On the other hand, there is no expanding scenario due to the shape of the potential, which is one of main differences from the previous results.

An alternative choice of the exterior metric is to take a Minkowski spacetime, which definitely describes a LDV solution. In this case, 
the collapsing behavior relies on the choice of the initial velocity of the edge, which yields two parallel collapse scenarios - contracting and expanding clouds. Provided it contracts, the clouds will not form an event horizon and one might expect an emergence of a naked singularity as an outcome of the collapse. But, the curvature singularity at ${\mathcal R}=0$ appears in infinite collapse time and the comoving observers will not experience it forever in the string frame. However, an appropriate physical interpretation should be addressed in Einstein frame, notifying an appearance of a naked singularity within finite time. In this sense, the collapse scenario includes a violation of the cosmic censorship hypothesis, which coincides with the results in refs. \cite{mr,mo}.

The analysis in this work has some limitation, despite of its simplicity of the calculations. For example, in our context, the anti-symmetric NS-field vanishes as noted before. The study on the formation of the black string with the NS-field might be achieved by assuming the inhomogeneity of the clouds as a source of the charge or the clouds of real charged particles in the low energy effective string theory.
\vspace{0.8cm}

\textbf{Acknowledgments}

The authors are grateful to the referee of the journal for the crucial indications and the fruitful suggestions. J. J. Oh would like to thank R. B. Mann for invaluable comments and M. I. Park, H. Shin, and E. J. Son for helpful discussions.
The work of S. Hyun was supported by
the Korea Research Foundation Grant KRF-2005-070-C00030. W. Kim was supported by the Science Research Center Program of the Korea Science and Engineering Foundation through the Center for Quantum Spacetime of Sogang University with grant number R11-2005-021. J. Jeong and J. J. Oh were supported by the Brain Korea 21(BK21) project funded by the Ministry of Education and Human Resources of Korea Government. 
 
\nc{\PR}[3]{Phys. Rev. {\bf #1} #2 (#3)}
\nc{\NPB}[3]{Nucl. Phys. {\bf B#1} #2 (#3)}
\nc{\PLB}[3]{Phys. Lett. {\bf B#1} #2 (#3)}
\nc{\PRD}[3]{Phys. Rev. {\bf D#1} #2 (#3)}
\nc{\PRL}[3]{Phys. Rev. Lett. {\bf #1} #2 (#3)}
\nc{\PREP}[3]{Phys. Rep. {\bf #1} #2 (#3)}
\nc{\EPJ}[3]{Eur. Phys. J. {\bf #1} #2 (#3)}
\nc{\PTP}[3]{Prog. Theor. Phys. {\bf #1} #2 (#3)}
\nc{\CMP}[3]{Comm. Math. Phys. {\bf #1} #2 (#3)}
\nc{\MPLA}[3]{Mod. Phys. Lett. {\bf A #1} #2 (#3)}
\nc{\CQG}[3]{Class. Quant. Grav. {\bf #1} #2 (#3)}
\nc{\NCB}[3]{Nuovo Cimento {\bf B#1} #2 (#3)}
\nc{\ANNP}[3]{Ann. Phys. (N.Y.) {\bf #1} #2 (#3)}
\nc{\GRG}[3]{Gen. Rel. Grav. {\bf #1} #2 (#3)}
\nc{\MNRAS}[3]{Mon. Not. Roy. Astron. Soc. {\bf #1} #2 (#3)}
\nc{\JHEP}[3]{JHEP {\bf #1} #2 (#3)}
\nc{\JCAP}[3]{JCAP {\bf #1}, #2 {(#3)}}
\nc{\ATMP}[3]{Adv. Theor. Math. Phys. {\bf #1} #2 (#3)}
\nc{\AJP}[3]{Am. J. Phys. {\bf #1} #2 (#3)}
\nc{\ibid}[3]{{\it ibid.} {\bf #1} #2 (#3)}
\nc{\ZP}[3]{Z. Physik {\bf #1} #2 (#3)}
\nc{\PRSL}[3]{Proc. Roy. Soc. Lond. {\bf A#1} #2 (#3)}
\nc{\LMP}[3]{Lett. Math. Phys. {\bf #1} #2 (#3)}
\nc{\AM}[3]{Ann. Math. {\bf #1} #2 (#3)}
\nc{\hepth}[1]{{\tt [hep-th/{#1}]}}
\nc{\grqc}[1]{{\tt [gr-qc/{#1}]}}
\nc{\astro}[1]{{\tt [astro-ph/{#1}]}}
\nc{\hepph}[1]{{\tt [hep-ph/{#1}]}}
\nc{\phys}[1]{{\tt [physics/{#1}]}}

\end{document}